\documentclass{amsart}
\usepackage{graphicx}
\newtheorem{thm}{Theorem}[section]

\newtheorem{lem}[thm]{Lemma}

\theoremstyle{definition}
\newtheorem{defn}[thm]{Definition}
\theoremstyle{remark}
\newtheorem{rem}[thm]{Remark}
\numberwithin{equation}{section}

\begin{document}

\title[Low Walsh Spectra]{Two Boolean functions with five-valued Walsh spectra and high nonlinearity}%
\author{Xiwang Cao, Lei Hu}%
\address{Xiwang Cao is with the School of Mathematical Sciences, Nanjing University of
Aeronautics and Astronautics, Nanjing 210016, China, email: {\tt
xwcao@nuaa.edu.cn},
Lei Hu is with the State Key Laboratory of Information security, Institute of Information Engineering, Chinese Academy of Sciences, Beijing 100093, China.
email:{\tt hu@is.ac.cn}}%

\begin{abstract}
For cryptographic systems the method of confusion and diffusion is used as a fundamental technique to achieve security. Confusion is reflected
in nonlinearity of certain Boolean functions describing the cryptographic transformation. In this paper, we present two balanced boolean functions which have low Walsh spectra and high nonlinearity. In the proof of the nonlinearity, a new method for evaluating some exponential sums over finite fields was provided.
\end{abstract}
\maketitle

\section{Introduction}

Boolean functions have wide applications in cryptography and coding theory. Let $\mathbb{F}_{2^n}$ be the finite field of $2^n$ elements and $f : \mathbb{F}_{2^n}\mapsto \mathbb{F}_2$ be a
Boolean function. For every nonnegative integer $r\leq n$, we denote by $nl_r(f)$ the minimum Hamming distance of $f$ and all
functions of algebraic degrees at most $r$ (in the case of $r=1$, we shall simply write $nl(f)$). In other words, $nl_r(f)$ equals the
distance from $f$ in its truth table representation to the Reed-Muller code $RM(r; n)$ of length $2^n$ and of order $r$. This distance
is called the $r$-th order nonlinearity of $f$. In application,
a Boolean function should have a high nonlinearity to ensure the cipher to defense linear approximation attack \cite{cai,carlet,matsui}
.

The conception of bent functions was defined by Rothaus in 1976
\cite{rothaus}. In words of coding theory, bent functions are those ones whose Hamming distance to the Reed-Muller code of order 1
equal to $2^{n-1}-2^{n/2-1}$ (where the number $n$ of variables should be
even), this distance is the maximum possible distance for a boolean
function can obtain. Thus, bent functions have the best nonlinearity, therefore, such
functions have been widely studied and have received a lot
of attention in the literature on cryptography,
coding theory, communication, and sequence design. Bent functions also have some combinatorial properties which can be applied in experimental design etc., see for example, \cite{carlet,cusick} and the references therein. However, the number of variables of a bent function should be even, and a bent function can not be balanced. These shortcomings restrict the application of bent functions.

In order to overcome these shortcomings, Chee, Lee and Kim \cite{Chee} introduced semi-bent functions at Asiacrypt' 94, those functions have no longer restrictions on the number of variables. Like bent functions, semi-bent functions are also widely studied in sequences and in cryptography (e.g., in the design of stream ciphers) \cite{Zeng2}. It is known that semi-bent functions can be balanced and have maximal nonlinearity among balanced plateaued functions and can possess desirable
properties such as low autocorrelation, propagation criteria,
resiliency, and high algebraic degree etc. \cite{mes}, \cite{Zheng1}, \cite{Zheng2}. However, there is only a few known constructions of semi-bent functions, almost all of them are derived from the trace function of a power polynomials, i.e., ${\rm Tr}(x^d)$,
for a suitably chosen $d$, see \cite{char,mes,zeng,Zeng2}. In view of this situation, Boolean functions with low Walsh spectra have their own meaning.

In the study of bent functions and semi-bent funxtions, two kinds of functions are often involved, namely, $f_1(x)={\rm Tr}(\lambda x^{2^m+1})$ and $f_2(x)={\rm Tr}(x^{(2^m-1)s})$, where ${\rm Tr}$ is the trace function from the finite field $\mathbb{F}_{2^{2m}}$ to $\mathbb{F}_2$.
In this note, we present a construction of Boolean functions which have low  Walsh spectra using an ``interleaving" technique. In an explicit expression, our constructions are
\begin{eqnarray}\label{f-1}
    &&f_{\lambda, \mu}(x)={\rm Tr}(\lambda x^{2^m+1})+{\rm Tr}(x){\rm Tr}(\mu x^{2^m-1}), \mbox{ and}\\
    &&g_{\lambda,\mu}(x)=(1+{\rm Tr}(x)){\rm Tr}(\lambda x^{2^m+1})+{\rm Tr}(x){\rm Tr}(\mu x^{2^m-1}),
\end{eqnarray}
where $\mu \in \mathbb{F}_{2^m}^*$  and $\lambda$ is a element in $\mathbb{F}_{2^{2m}}$ satisfying $\lambda+\lambda^{2^m}=1$. The main contribution of this note is to calculate the Walsh spectra of the functions in (1.1) and (1.2). We will show that the Walsh spectra of $f_{\lambda, \mu}$ (resp. $g_{\lambda,\mu}$) is at most five-valued and each value is divided by $2^{m}$, and the nonlinearity is bigger than or equal to $2^{n-1}-3\cdot2^{m-1}$ (resp. $2^{n-1}-2\cdot2^{m-1}$). This nonlinearity is close to the possible maximum value $2^{n-1}-2^{m-1}$. When $m$ is odd, the function $g_{\lambda,\mu}$ is balanced. Comparing with bent functions, the nonlinearity of those functions in forms (1.1) and (1.2) are near optimal, and they can be changed to balanced function by adding a linear function while bent functions can not. Moreover, these two functions have higher algebraic degree than that of those related bent functions. In this sense, we proposed a new method for constructing new balanced boolean functions with low Walsh spectra and high nonlinearity. In addition, in this proof of the nonlinearity, we provide a new method for computing some exponential sums over finite fields, the explicit value of certain exponential sums were obtained.

\section{Notations and Preliminaries}

\subsection{ Trace representations of Boolean functions}

Let $n$ be a positive integer and $\mathbb{F}_{2^n}$ be the finite field with $2^n$ elements. A Boolean function on $\mathbb{F}_{2^n}$ is
an $\{0,1\}$-valued function from $\mathbb{F}_{2^n}$ to $\mathbb{F}_2$.

For any positive integer $n$, and for any positive integer $k$ dividing $n$, the trace
function from $\mathbb{F}_{2^n}$ to $\mathbb{F}_{2^k}$, denoted by ${\rm Tr}_k^n$, is the mapping defined as
\begin{equation*}
   {\rm Tr}_k^n(x)=x+x^{2^k}+x^{2^{2k}}+\cdots+x^{2^{n-k}}.
\end{equation*}
In particular, the absolute trace over $\mathbb{F}_2$ is the function
${\rm Tr}_1^n(x)=\sum_{i=0}^{n-1}x^{2^i}$ for $k=1$. Recall that, for every integer $k$ dividing
$n$, the trace function satisfies the transitivity property, that
is, for all $x\in \mathbb{F}_{2^n}$, it holds that \cite{lidl}
\begin{equation*}
    {\rm Tr}_1^n(x)={\rm Tr}_1^k({\rm Tr}_k^n(x)).
\end{equation*}
It is known \cite{carlet, mes} that every nonzero Boolean function $g$ defined on $\mathbb{F}_{2^n}$ has a
unique trace expansion of the form
\begin{equation*}
    g(x)=\sum_{j\in \Gamma_n}{\rm Tr}_1^{o(j)}(ax^j)+\epsilon(1+x^{2^n-1})
\end{equation*}
where $\Gamma_n$ is the set of integers obtained by choosing one element in each cyclotomic coset of $2$ modulo $(2^n-1)$ and $o(j)$ is the size of the cyclotomic coset of $2$ modulo $(2^n-1)$  containing $j$, $a_j\in \mathbb{F}_{2^{o(j)}}$, and $\epsilon=wt(g)$ modulo $2$ where $wt(g)$ is the Hamming weight
of the image vector of $g$, that is, the cardinality of its support set
\begin{equation*}
Supp(g):=\{x\in \mathbb{F}_{2^n}|g(x)=1\}.
\end{equation*}

For every positive integer $r$, let $r=\sum_{j}r_j2^j$ be the $2$-adic representation of $r$. Denote $wt(r)=\sum_jr_j$. For every polynomial $f(x)=\sum_{i=0}^{n}a_ix^i\in \mathbb{F}_q[x]$, the number
\begin{equation*}
   d:=\max\{wt(i): a_i\neq 0\}
\end{equation*}
is called the {\it algebraic degree} of $f$. In order to defense algebraic attack on some cryptography protocol, functions with high algebraic degree are desirable. See \cite{carlet1, carlet} for details.

\subsection{ Walsh transform and nonlinearities of Boolean functions, bent-functions, semi-bent functions}

Let $f$ be a Boolean function from $\mathbb{F}_{2^n}$ to $\mathbb{F}_2$.

{\defn For every element $a\in \mathbb{F}_{2^n}$, the {\it Walsh (Hadamard) Transformation} of $f$ at the point $a$ is defined by
\begin{equation*}
    W_f(a)=\sum_{x\in \mathbb{F}_{2^n}}(-1)^{f(x)+{\rm Tr}_1^n(ax)}.
\end{equation*}}

The $2^n$-dimensional vector $\mathfrak{F}=(f(0),f(\gamma^0),\cdots,f(\gamma^{2^n-2}))$ is called the truth table representation of $f$, where $\gamma$ is a fixed primitive element of $\mathbb{F}_{2^n}$. Set $0=\gamma^{-\infty}$. The Hamming distance of two Boolean functions $f$ and $g$, denoted by $d(f,g)$, is defined by the number of distinct corresponding coordinates of the truth table representation of these functions, i.e.,
\begin{equation*}
    d(f,g)=|\{i: f(\gamma^i)\neq g(\gamma^i), i=-\infty, i=1,\cdots, 2^n-2\}|.
\end{equation*}
The $r$-th nonlinearity of $f$ is defined by the minimum Hamming distance of $f$ and all the Boolean functions of degree less than or equal to $r$, i.e.,
\begin{equation*}
   nl_r(f)=\underset{g}{ \min }\{d(f,g): 1\leq \deg(g)\leq r\}.
\end{equation*}
When $r=1$, it deduces the nonlinearity of $f$ as
\begin{equation*}
   nl(f)=2^{n-1}-\frac{1}{2}\underset{a\in \mathbb{F}_{2^n}}{\max}|W_f(a)|.
\end{equation*}
The possible maximum value of the nonlinearity of a Boolean function is $2^{n-1}-2^{n/2-1}$, see \cite{carlet1} for details.

{\defn A Boolean function $f$ from $\mathbb{F}_{2^n}$ to $\mathbb{F}_2$ is called bent if $W_f(a)=\pm 2^{\frac{n}{2}}$ for all $a\in \mathbb{F}_{2^n}$, or is called semi-bent if
$W_f(a)\in \{0, \pm 2^{\lfloor\frac{n+1}{2}\rfloor\}}$ for all $a\in \mathbb{F}_{2^n}$. A function $f$ from $\mathbb{F}_{p^n}$ to $\mathbb{F}_p$ is called {\it plateaued} if $W_f(a)=
A $ or $0$ for all $a\in \mathbb{F}_{p^n}$, where $A$ is a constant and $p$ is a prime.}


It is well known that the maximal algebraic degree of a
binary bent function from $\mathbb{F}_{2^{2m}}$ to $\mathbb{F}_2$ is $m$. (see \cite{rothaus}).
\subsection{Polar decomposition}

Let $n=2m$. Denote the subgroup of $(2^m+1)$-th roots of unity in $\mathbb{F}_{2^n}$ by $\mathfrak{S}$, i.e., $\mathfrak{S}=\{z\in \mathbb{F}_{2^n}| z^{2^m+1}=1\}$. For every $x\in \mathbb{F}_{2^n}^*=\mathbb{F}_q\setminus \{0\}$, there is a unique polar decomposition of $x$ as $x=yz$ where $y\in \mathbb{F}_{2^m}^*$ and $z\in \mathfrak{S}$. In fact, $y=x^{(2^m+1)2^{m-1}}, z=x^{(2^m-1)2^{m-1}}$. Denote $x^{2^m}$ by $\overline{x}$. Then for every $x\in \mathbb{F}_{2^n}^*$, $x\in \mathbb{F}_{2^m}$ if and only if $x=\overline{x}$, and $x\in \mathfrak{S}$ if and only if $\overline{x}=x^{-1}$. it is evident that for every $x\in \mathbb{F}_{2^n}^*$, one has that $x+\overline{x}, x\overline{x}\in \mathbb{F}_{2^m}$ and $x/\overline{x},\overline{x}/x\in \mathfrak{S}$. Note that $x\mapsto \overline{x}$ is an isomorphism of the finite field $\mathbb{F}_{2^n}$.

\subsection{ Kloosterman sums}

For every $a,b\in \mathbb{F}_{2^m}$, Kloosterman sum is defined by
\begin{eqnarray}
  \label{equ-k}k_m(a,b) &=& \sum_{x\in \mathbf{F}_{2^m}^*}(-1)^{{\rm Tr}_m(ax+bx^{-1})}.
 \end{eqnarray}
 It is easy to check that $k_m(a,b)=k_m(ab,1)=k_m(1,ab)$. For simplicity, denote $k_m(a,1)=k_m(1,a)$ by $k_m(a)$. Moreover, The Kloosterman sum $k_m(a,b)$ can be calculated recursively, that is,
 if we define
 \begin{equation*}
    k_m^{(s)}(a)=\sum_{\gamma\in \mathbb{F}_{{2^{ms}}}^*}\chi^{(s)}(a\gamma+\gamma^{-1}), a\in \mathbb{F}_{2^m},
 \end{equation*}
where $\chi^{(s)}$ is the lifting of $\chi(x)=(-1)^{{\rm Tr}(x)}$ to $\mathbb{F}_{2^{ms}}$, then
 \begin{equation}\label{f-41}
    k_m^{(s)}(a)=-k_{m}^{(s-1)}(a)k_m^{(1)}(a)-2^mk_{m}^{(s-2)}(a),
 \end{equation}
 where we put $k_m^{(0)}(a,b)=-2$ and $k_m^{(1)}(a)=k(a)$. Moreover, for all $a,b\in \mathbb{F}_{2^m}^*$, one has that
  \begin{equation*}
   |k_m(a,b)|\leq 2\sqrt{2^m},
\end{equation*}
See \cite{lidl} for details.


Note also that the values of Kloosterman sums over $\mathbb{F}_{2^m}$
were determined by Lachaud and Wolfmann in \cite{lachaud}.

\begin{lem}\label{lem-43} (\cite{lachaud}) The set $\{k_m(\lambda),\lambda\in \mathbb{F}_{2^m}\}$ is the set of all the integers $s\equiv -1({\rm mod}\ 4)$ in the range
\begin{equation*}
    \left[-2^{\frac{m}{2}+1},2^{\frac{m}{2}+1}\right].
\end{equation*}
\end{lem}

\section{Main results}

In this section, we always assume that $n=2m$ is an even positive integer. We fix the notations as in the previous section.

Before we go to present our main results on the Walsh spectra of the functions $f_{\lambda.\mu}$ and $g_{\lambda,\mu}$, we quote a lemma which will be used frequently in the sequel.

{\lem \label{lem-31}(\cite{dobbertin})  Let $H_1=\{x\in \mathbb{F}_{2^m}^*|{\rm Tr}_1^m(1/x)=1\}$. Then
\begin{equation}\label{f-7}
    H_1=\{u+\overline{u}|u\in \mathfrak{S}\setminus\{1\}\}.
\end{equation}
Moreover, the mapping $u\mapsto {\rm Tr}^n_m(u)=u+\overline{u}$ from $\mathfrak{S}\setminus \{1\}$
to $H_1$ is onto and 2-to-1, and ${\rm Tr}^n_m(u)={\rm Tr}^n_m(v)$ if and only
if $u=v$ or $u=\overline{v}$ for $u,v\in \mathfrak{S}\setminus \{1\}$. As a consequence, for every element $a\in \mathbb{F}_{2^m}^*$, the equation
\begin{equation}\label{f-31}
    X+X^{-1}=1/a
\end{equation}
has two distinct roots in $\mathfrak{S}$ if ${\rm Tr}^m_1(a)=1$ and no root in $\mathfrak{S}$ otherwise.
}

\subsection{Walsh spectra of $f_{\lambda.\mu}$}

In this subsection, we prove the following main result of this note.

{\thm \label{thm-31} 
Define a Boolean function \begin{equation}\label{f-1}
    f_{\lambda,\mu}(x)={\rm Tr}_1^n(\lambda x^{2^m+1})+{\rm Tr}_1^n(x){\rm Tr}_1^n(\mu x^{2^m-1}),
\end{equation}
where $\lambda$ is an element in $\mathbb{F}_{2^n}$ satisfying $\lambda+\overline{\lambda}=1$ and $\mu \in \mathbb{F}_{2^m}^*$. Then the Walsh spectra of $f_{\lambda,\mu}$ is contained in the set
\begin{equation*}
 \{0, \pm 2^m, 2^{m+1}, 3\cdot 2^m\}.
\end{equation*}
As a consequence, the nonlinearity of $f$ is
\begin{equation*}
    nl(f)\geq 2^{n-1}-3\cdot2^{m-1}.
\end{equation*}
}

Note that the algebraic degree of $f_{\lambda,\mu}$ is $m+1$ and thus it can not be bent. However, from Theorem \ref{thm-31}, the function $f_{\lambda,\mu}$ still has a relatively high nonlinearity.

\begin{proof}
Define two sets $T_0$ and $ T_1$ as
\begin{equation*}
   T_i=\{x\in \mathbb{F}_{2^n}|{\rm Tr}(x)=i\}, i=0,1.
\end{equation*}
It is obvious that for $a\neq 0$,
\begin{eqnarray*}
  W_f(a)&= &\sum_{x\in \mathbb{F}_{2^n}}(-1)^{f(x)+{\rm Tr}_1^n(ax)} \\
  &= &\sum_{x\in T_0}(-1)^{{\rm Tr}_1^n(\lambda x^{2^m+1}+ax)}+\sum_{x\in T_1}(-1)^{{\rm Tr}_1^n(\lambda x^{2^m+1}+\mu x^{2^m-1}+ax)}\\
  &:=&S_0(a)+S_1(a),
\end{eqnarray*}
where $S_0(a)$ and $S_1(a)$ denote the two summations.
Since
\begin{eqnarray*}
  & &\sum\limits_{x\in T_1}(-1)^{{\rm Tr}_1^n(\lambda x^{2^m+1}+(a+1)x)} \\
 &=& \sum\limits_{x\in T_1}(-1)^{{\rm Tr}_1^n(\lambda x^{2^m+1}+ax)}(-1)^{{\rm Tr}_1^n(x)}
 =-\sum\limits_{x\in T_1}(-1)^{{\rm Tr}_1^n(\lambda x^{2^m+1}+ax)}
\end{eqnarray*}
and
$$
\sum_{x\in T_0}(-1)^{{\rm Tr}_1^n(\lambda x^{2^m+1}+(a+1)x)}=\sum_{x\in T_0}(-1)^{{\rm Tr}_1^n(\lambda x^{2^m+1}+ax)},
$$
we have
\begin{eqnarray*}
    2S_0(a)&=&\sum_{x\in \mathbb{F}_{2^n}}(-1)^{{\rm Tr}_1^n(\lambda x^{2^m+1}+(a+1)x)}+\sum_{x\in \mathbb{F}_{2^n}}(-1)^{{\rm Tr}_1^n(\lambda x^{2^m+1}+ax)}\\
    &=&2+\sum_{x\in \mathbb{F}_{2^n}^*}(-1)^{{\rm Tr}_1^n(\lambda x^{2^m+1}+(a+1)x)}+\sum_{x\in \mathbb{F}_{2^n}^*}(-1)^{{\rm Tr}_1^n(\lambda x^{2^m+1}+ax)}.
\end{eqnarray*}
By the condition that $\lambda+\overline{\lambda}=1$, it follows that
$${\rm Tr}_1^n(\lambda x^{2^m+1})={\rm Tr}_1^m(x\overline{x}(\lambda+\overline{\lambda}))={\rm Tr}_1^m(x\overline{x}).$$
Using the polar decomposition, we have
\begin{eqnarray*}
  \sum_{x\in \mathbb{F}_{2^n}^*}(-1)^{{\rm Tr}_1^n(\lambda x^{2^m+1}+ax)}&=& \sum_{y\in \mathbb{F}_{2^m}^*,z\in \mathfrak{S}}(-1)^{{\rm Tr}_1^m(y(1+az+\overline{a}z^{-1})}.
\end{eqnarray*}
Let $a=a_0a_1$ be the polar decomposition of $a$. In fact, $a_0=(\overline{a}a)^{2^{m-1}}, a_1=(\overline{a}/a)^{2^{m-1}}$. Then
\begin{equation*}
    \sum_{y\in \mathbb{F}_{2^m}^*,z\in \mathfrak{S}}(-1)^{{\rm Tr}_1^m(y(1+a_0(a_1z)+a_0(a_1z)^{-1})}= \sum_{y\in \mathbb{F}_{2^m}^*,z\in \mathfrak{S}}(-1)^{{\rm Tr}_1^m(y(1+a_0z+a_0z^{-1})}.
\end{equation*}

By Lemma \ref{lem-31} and the fact that $\sum\limits_{y\in \mathbb{F}_{2^m}^*}(-1)^{{\rm Tr}_1^m(by)}=-1$ for any $b\in  \mathbb{F}_{2^m}^*$, if ${\rm Tr}_1^m(a\overline{a})={\rm Tr}_1^m(a_0)=0$, then
\begin{eqnarray*}
   & & \sum_{y\in \mathbb{F}_{2^m}^*,z\in \mathfrak{S}}(-1)^{{\rm Tr}_1^m(y(1+a_0z+a_0z^{-1})}  \\
   &=& \sum_{z\in \mathfrak{S}}\sum_{y\in \mathbb{F}_{2^m}^*}(-1)^{{\rm Tr}_1^m(y(1+a_0z+a_0z^{-1})}
=\sum_{z\in \mathfrak{S}}(-1)=-2^m-1,
\end{eqnarray*}
while if  ${\rm Tr}_1^m(a\overline{a})={\rm Tr}_1^m(a_0)=1$, then
\begin{eqnarray*}
   & & \sum_{y\in \mathbb{F}_{2^m}^*,z\in \mathfrak{S}}(-1)^{{\rm Tr}_1^m(y(1+a_0z+a_0z^{-1})}  \\
   &=& \sum_{z\in \mathfrak{S}}\sum_{y\in \mathbb{F}_{2^m}^*}(-1)^{{\rm Tr}_1^m(y(1+a_0z+a_0z^{-1})}
=2(2^m-1)+(2^m-1)(-1)=2^m-1,
\end{eqnarray*}
thus, we have
\begin{equation*}
    \sum_{x\in \mathbb{F}_{2^n}^*}(-1)^{{\rm Tr}_1^n(\lambda x^{2^m+1}+ax)}=\left\{\begin{array}{ll}
                                                                                 -2^m-1,& \mbox{ if ${\rm Tr}_1^m(a\overline{a})=0$}, \\
                                                                                 2^m-1, & \mbox{ if ${\rm Tr}_1^m(a\overline{a})=1$.}
                                                                               \end{array}
    \right.
\end{equation*}
Similarly, we have
\begin{equation*}
    \sum_{x\in \mathbb{F}_{2^n}^*}(-1)^{{\rm Tr}_1^n(\lambda x^{2^m+1}+(a+1)x)}=\left\{\begin{array}{ll}
                                                                                 -2^m-1,& \mbox{ if ${\rm Tr}_1^m((a+1)(\overline{a}+1))=0$}, \\
                                                                                 2^m-1, & \mbox{ if ${\rm Tr}_1^m((a+1)(\overline{a}+1))=1$.}
                                                                               \end{array}
    \right.
\end{equation*}
Since ${\rm Tr}_1^m((a+1)(\overline{a}+1))=m+{\rm Tr}_1^m(a\overline{a})+{\rm Tr}_1^n(a)$, we have
\begin{equation}\label{f-3}
    S_0(a)=\left\{\begin{array}{ll}
                -2^{m} ,& \mbox{if ${\rm Tr}_1^n(a)=m(\mathrm{mod} \,2)$ and ${\rm Tr}_1^m(a\overline{a})=0$}, \\
                 2^m,& \mbox{if ${\rm Tr}_1^n(a)=m(\mathrm{mod} \,2)$ and ${\rm Tr}_1^m(a\overline{a})=1$}, \\
                 0,& \mbox{if ${\rm Tr}_1^n(a)\not=m(\mathrm{mod} \,2)$}.
               \end{array}
    \right.
\end{equation}
%
Using a similar procedure as above, we have
\begin{equation*}
    2S_1(a)=\sum_{x\in \mathbb{F}_{2^n}}(-1)^{{\rm Tr}_1^n(\lambda x^{2^m+1}+\mu x^{2^m-1}+ax)}-\sum_{x\in \mathbb{F}_{2^n}}(-1)^{{\rm Tr}_1^n(\lambda x^{2^m+1}+\mu x^{2^m-1}+(a+1)x)}.
\end{equation*}
For the first summation, we have
\begin{eqnarray*}
  && \sum_{x\in \mathbb{F}_{2^n}}(-1)^{{\rm Tr}_1^n(\lambda x^{2^m+1}+\mu x^{2^m-1}+ax)} \\
   &=& 1+\sum_{z\in \mathfrak{S}}(-1)^{{\rm Tr}_1^m(\mu' z+\mu' z^{-1})}\sum_{y\in \mathbb{F}_{2^m}^*}(-1)^{{\rm Tr}_1^m(y(1+az+\overline{a}z^{-1}))}
\end{eqnarray*}
where $\mu'=\mu^{1/2}$. In what follows, we still use $\mu$ instead of $\mu'$. This does not  matter since $k_m(\mu)=k_m(\mu^2)$.

If ${\rm Tr}^m_1(a\overline{a})=1$, then $1+az+\overline{a}z^{-1}=0$ has two distinct roots $\theta_1$ and $\theta_2$ in $\mathfrak{S}$. In this case, we have
\begin{eqnarray*}
  && \sum_{x\in \mathbb{F}_{2^n}}(-1)^{{\rm Tr}_1^n(\lambda x^{2^m+1}+\mu x^{2^m-1}+ax)} \\
   &=& 1+\sum_{z\in \mathfrak{S}}(-1)^{{\rm Tr}_1^m(\mu(z+z^{-1}))}\sum_{y\in \mathbb{F}_{2^m}^*}(-1)^{{\rm Tr}_1^m(y(1+az+\overline{a}z^{-1}))}\\
   &=&1+[(-1)^{{\rm Tr}^m_1(\mu(\theta_1+\overline{\theta_1}))}+(-1)^{{\rm Tr}^m_1(\mu(\theta_2+\overline{\theta_2}))}](2^m-1)\\
   &&+\sum_{z\in \mathfrak{S}\setminus\{\theta_1,\theta_2\}}(-1)^{{\rm Tr}_1^m(\mu(z+z^{-1}))}(-1)\\
   &=&1+2^{m}A-\sum_{z\in \mathfrak{S}}(-1)^{{\rm Tr}_1^m(\mu(z+z^{-1}))},
\end{eqnarray*}
where $A=(-1)^{{\rm Tr}^m_1(\mu(\theta_1+\overline{\theta_1}))}+(-1)^{{\rm Tr}^m_1(\mu(\theta_2+\overline{\theta_2}))}=(-1)^{{\rm Tr}^n_1(\mu\theta_1)}+(-1)^{{\rm Tr}^n_1(\mu\theta_2)}.$

It is known that for $\mu\neq 0$,
\begin{equation}\label{f-kl}
   \sum_{z\in \mathfrak{S}}(-1)^{{\rm Tr}_1^m(\mu(z+z^{-1}))}=-k_m(\mu).
\end{equation}
where $k_m(\mu)$ is the Kloosterman sum. See for example \cite{char,jhk}.

If  ${\rm Tr}^m_1(a\overline{a})=0$, then $1+a_0z+a_0z^{-1}$ has no root in $\mathfrak{S}$. In this case, we have
\begin{equation}\label{f-9}
    \sum_{x\in \mathbb{F}_{2^n}}(-1)^{{\rm Tr}_1^n(\lambda x^{2^m+1}+\mu x^{2^m-1}+ax)}=1+(-1)\sum_{z\in \mathfrak{S}}(-1)^{{\rm Tr}(\mu(z+z^{-1}))}=1+k_m({\mu}).
\end{equation}
Therefore, we have
\begin{eqnarray}\label{f-11}
  &&\sum_{x\in \mathbb{F}_{2^n}}(-1)^{{\rm Tr}_1^n(\lambda x^{2^m+1}+\mu x^{2^m-1}+ax)}\\
 \nonumber &=&\left\{\begin{array}{ll}
                                                                                      1+k_m({\mu}),&\mbox{if  ${\rm Tr}^m_1(a\overline{a})=0$}, \\
                                                                                       2^{m}A+1+k_m(\mu), & \mbox{if  ${\rm Tr}^m_1(a\overline{a})=1$},
                                                                                     \end{array}
  \right.
\end{eqnarray}

Similarly, we have
\begin{eqnarray}\label{f-13}
  &&\sum_{x\in \mathbb{F}_{2^n}}(-1)^{{\rm Tr}_1^n(\lambda x^{2^m+1}+\mu x^{2^m-1}+(a+1)x)}\\
 \nonumber &=&\left\{\begin{array}{cc}
                                                                                      1+k_m({\mu}),&\mbox{if  ${\rm Tr}^m_1((a+1)(\overline{a}+1))=0$}, \\
                                                                                       2^{m}B+1+k_m(\mu), & \mbox{if  ${\rm Tr}^m_1((a+1)(\overline{a}+1))=1$}.
                                                                                     \end{array}
  \right.
\end{eqnarray}
where $B=(-1)^{{\rm Tr}^m_1(\mu({\tau_1}+{\tau_1}^{-1}))}+(-1)^{{\rm Tr}^m_1(\mu(\tau_2+\tau_2^{-1}))}=(-1)^{{\rm Tr}^n_1(\mu{\tau_1})}+(-1)^{{\rm Tr}^n_1(\mu\tau_2)}$, $\tau_i\in \mathfrak{S}, i=1,2 $ are the roots of $1+(a+1)X+(\overline{a}+1)X^{-1}=0$.
In a summary, we have

\begin{equation}\label{f-15}
    S_1(a)=\left\{\begin{array}{ll}
               0 ,& \mbox{if ${\rm Tr}_1^n(a)=m(\mathrm{mod} \,2)$ and ${\rm Tr}_1^m(a\overline{a})=0$}, \\
                 -2^{m-1}B,& \mbox{if ${\rm Tr}_1^n(a)\not=m(\mathrm{mod} \,2)$ and ${\rm Tr}_1^m(a\overline{a})=0$}, \\
                 2^{m-1}(A-B) ,& \mbox{if ${\rm Tr}_1^n(a)=m(\mathrm{mod} \,2)$ and ${\rm Tr}_1^m(a\overline{a})=1$},\\
               2^{m-1}A,& \mbox{if ${\rm Tr}_1^n(a)\not=m(\mathrm{mod} \,2)$ and ${\rm Tr}_1^m(a\overline{a})=1$}.
               \end{array}
    \right.
\end{equation}

Combining (\ref{f-3}) and  (\ref{f-15}), we obtain that

\begin{equation}\label{f-17}
    W_f(a)=\left\{\begin{array}{ll}
               -2^m ,& \mbox{if ${\rm Tr}_1^n(a)=m(\mathrm{mod} \,2)$ and ${\rm Tr}_1^m(a\overline{a})=0$}, \\
                 -2^{m-1}B,& \mbox{if ${\rm Tr}_1^n(a)\not=m(\mathrm{mod} \,2)$ and ${\rm Tr}_1^m(a\overline{a})=0$}, \\
                 2^{m-1}(A-B+2),& \mbox{if ${\rm Tr}_1^n(a)=m(\mathrm{mod} \,2)$ and ${\rm Tr}_1^m(a\overline{a})=1$},\\
                2^{m-1}A,& \mbox{if ${\rm Tr}_1^n(a)\not=m(\mathrm{mod} \,2)$ and ${\rm Tr}_1^m(a\overline{a})=1$}.
               \end{array}
    \right.
\end{equation}

For $W_f(0)$, we compute that
\begin{eqnarray*}
  W_f(0)&=& \sum_{x\in \mathbb{F}_{2^n}}(-1)^{f(x)}  \\
 &=&\sum_{x\in T_0}(-1)^{{\rm Tr}_1^n(\lambda x^{2^m+1})}+\sum_{x\in T_1}(-1)^{{\rm Tr}_1^n(\lambda x^{2^m+1}+\mu x^{2^m-1})}.
\end{eqnarray*}
For the first summation, we have
\begin{eqnarray*}
  &&2\cdot \sum_{x\in T_0}(-1)^{{\rm Tr}_1^n(\lambda x^{2^m+1})}\\
  &=& \sum_{x\in \mathbb{F}_{2^n}}(-1)^{{\rm Tr}_1^n(\lambda x^{2^m+1})}+\sum_{x\in \mathbb{F}_{2^n}}(-1)^{{\rm Tr}_1^n(\lambda x^{2^m+1}+x)} \\
   &=& 2+\sum_{y\in \mathbb{F}_{2^m}^*,z\in \mathfrak{S}}(-1)^{{\rm Tr}_1^m(y)}+\sum_{y\in \mathbb{F}_{2^m}^*,z\in \mathfrak{S}}(-1)^{{\rm Tr}_1^m(y(1+z+z^{-1}))}\\
   &=& 2+(-1)(2^m+1)+\sum_{y\in \mathbb{F}_{2^m}^*,z\in \mathfrak{S}}(-1)^{{\rm Tr}_1^m(y(1+z+z^{-1}))},
\end{eqnarray*}
which is equal to
$1-2^m+(2^m+1)(-1)=-2^{m+1}$ for even $m$ and equal to
$1-2^m+2(2^m-1)+(2^m-1)(-1)=0$ for odd $m$ by Lemma \ref{lem-31}.
For the second summation, we have
\begin{eqnarray*}
  && 2\cdot\sum_{x\in T_1}(-1)^{{\rm Tr}_1^n(\lambda x^{2^m+1}+\mu x^{2^m-1})}\\
  &=& \sum_{x\in \mathbb{F}_{2^n}}(-1)^{{\rm Tr}_1^n(\lambda x^{2^m+1}+\mu x^{2^m-1})}-\sum_{x\in \mathbb{F}_{2^n}}(-1)^{{\rm Tr}_1^n(\lambda x^{2^m+1}+\mu x^{2^m-1}+x)}  \\
   &=&\sum_{z\in \mathfrak{S}} (-1)^{{\rm Tr}_1^m(\mu(z+z^{-1}))} \Big(\sum_{y\in \mathbb{F}_{2^m}^*}(-1)^{{\rm Tr}_1^m(y)}-\sum_{y\in \mathbb{F}_{2^m}^*}(-1)^{{\rm Tr}_1^m(y(1+z+z^{-1}))}\Big),
\end{eqnarray*}
which again by Lemma \ref{lem-31} is equal to
$0$ for even $m$ and equal to
$$2(-1)^{{\rm Tr}_1^m(\mu(\rho_1+\rho_1^{-1}))}(-1-(2^m-1))=2^{m+1}(-1)^{{\rm Tr}_1^m(\mu(\rho_1+\rho_1^{-1}))}$$ for odd $m$, where $\rho_1$ and $\overline{\rho_1}$ are two distinct roots of $1+z+z^{-1}=0$ in $\mathfrak{S}$.
Thus, when $m$ is even, $W_f(0)=-2^m$, and when $m$ is odd, $W_f(0)=2^{m}(-1)^{{\rm Tr}_1^m(\mu(\rho_1+\rho_1^{-1}))}$.

By the fact that $A,B\in \{0,-2,2\}$, we know that the spectra of $f$ is contained in the set $\{0,\pm 2^m, 2^{m+1},3\cdot2^m\}$.

Denote by $N_i$ the number of  $a\in \mathbb{F}_{2^n}$ such that $W_f(a)=2^mi$, $i=-1,0,1,2,3$.
Then by the inverse of the Walsh transformation and Parseval's equality, we have
$$\left\{\begin{array}{l}
  N_0+N_1+N_{-1}+N_2+N_{3}=2^n, \\
  N_1-N_{-1}+2N_2+3N_{3}=2^m, \\
  N_1+N_{-1}+4N_2+9N_{3}=2^n,
\end{array}
\right.$$
which gives us
$$\left\{\begin{array}{l}
  N_0=3N_2+8N_{3}, \\
  N_1=2^{n-1}+2^{m-1}-3N_2-6N_{3}, \\
  N_{-1}=2^{n-1}-2^{m-1}-N_2-3N_{3}.
\end{array}
\right.$$

Below, we show that $N_0>0$ which also implies that $f_{\lambda,\mu}$ is not a bent function. We assume that $m$ is odd (the case of $m$ is even can be handled similarly). Since $\theta_1+\theta_2=1/a$, $\tau_1+\tau_2=1/(a+1)$, it follows from (3.11) that the set of those elements $a$ such that $W_f(a)=0$ contains the set
\begin{equation*}
    Q_1=\{a\in \mathbb{F}_{2^n}^*: {\rm Tr}_1^n(\mu /a)=1, {\rm Tr}_1^n(a)=0, {\rm Tr}_1^m(a\overline{a})=1\}
\end{equation*}
join the set
\begin{equation*}
    Q_2=\{a\in \mathbb{F}_{2^n}^*: {\rm Tr}_1^n(\mu /(a+1))=1, {\rm Tr}_1^n(a)=0, {\rm Tr}_1^m(a\overline{a})=0\}.
\end{equation*}
We consider the set
\begin{equation*}
   Q:=\{a\in \mathbb{F}_{2^n}^*: {\rm Tr}_1^n(\mu /a)=1, {\rm Tr}_1^n(\mu /(a+1))=1, {\rm Tr}_1^n(a)=0 \},
\end{equation*}
then it is obvious that $Q\subseteq Q_1\cup Q_2$, and
\begin{eqnarray*}
  4|Q|&=& \sum_{a\in \mathbb{F}_{2^n}-\{0,1\}}(1-(-1)^{{\rm Tr}_1^n(\mu /a)})(1-(-1)^{{\rm Tr}_1^n(\mu/(a+1))})(1+(-1)^{{\rm Tr}_1^n(a)}) \\
&=& \sum_{a\in \mathbb{F}_{2^n}-\{0,1\}}\left[1-(-1)^{{\rm Tr}_1^n(\mu /a)}-(-1)^{{\rm Tr}_1^n(\mu/(a+1))}+(-1)^{{\rm Tr}_1^n(\mu/(a^2+a))}\right. \\
&&\left.+(-1)^{{\rm Tr}_1^n(\mu a)}-(-1)^{1+{\rm Tr}_1^n(a+\mu/a)}-(-1)^{{\rm Tr}_1^n(a+\mu/(a+1))}+(-1)^{{\rm Tr}_1^n(a+\frac{\mu}{a^2+a})}\right]\\
 &=& 2^n-2k_n(\mu)+\sum_{a\in \mathbb{F}_{2^n}-\{0,1\}}(-1)^{{\rm Tr}_1^n(\mu /(a^2+a))}+\sum_{a\in \mathbb{F}_{2^n}-\{0,1\}}(-1)^{{\rm Tr}_1^n(a+\frac{\mu}{a^2+a})}
 \end{eqnarray*}
Now, we compute that
\begin{eqnarray*}
 && \sum_{a\in \mathbb{F}_{2^n}-\{0,1\}}(-1)^{{\rm Tr}_1^n(\mu (\frac{1}{a^2+a}))}\\
   &=&2\sum_{a\in \mathbb{F}_{2^n}^*, {\rm Tr}_1^n(a)=0}(-1)^{{\rm Tr}_1^n(\mu /a)} \\
   &=&\sum_{a\in \mathbb{F}_{2^n}^*}(-1)^{{\rm Tr}_1^n(\mu /a)}(1+(-1)^{{\rm Tr}_1^n(a)})\\
   &=&-1+k_n(\mu).
\end{eqnarray*}
Therefore, one has that
\begin{equation*}
  4|Q|=2^n-1-k_n(\mu)+\sum_{a\in \mathbb{F}_{2^n}-\{0,1\}}(-1)^{{\rm Tr}_1^n(a+\frac{\mu}{a^2+a})}.
\end{equation*}
By \cite{momo}, Theorem 2, we know that
\begin{equation*}
    |\sum_{a\in \mathbb{F}_{2^n}-\{0,1\}}(-1)^{{\rm Tr}_1^n(a+\frac{\mu}{a^2+a})}|\leq 4\cdot 2^m.
\end{equation*}
Thus we have from Lemma \ref{lem-43} that
\begin{equation*}
   4|Q|\geq 2^n-2^m-2^{m+2}=2^m(2^m-5)>0
   \end{equation*}
   when $m>2$. This proves that $N_0>0$.
\end{proof}

{\rem (1) It is known that in the applications of communication and sequences design, Boolean functions with lower or less Walsh spectra are desirable. The function in Theorem 3.2 is at most five valued and the nonlinearity is relatively big. 

(2) Interestingly, we find that the Walsh spectra of $f_{\lambda,\mu}$ is the same as that of the function ${\rm Tr}_1^{2m}(x^{2^m+3})$, see \cite{helleseth2}.

(3) Taking $\mu=1$ and using MAGMA, we obtain the Walsh spectra and the frequency of the function $f_{\lambda,1}$ in the following table.

\begin{center}

$m=4$\hskip 2 cm $m=5$\hskip 2 cm $m=6$
\begin{tabular}
{|c|c|c|c|c|c|}
  \hline
  $W_{f,1}(a)$ & frequency &  $W_{f,1}(a)$  & frequency & $W_{f,1}(a)$   & frequency \\
 \hline
  0& 80 & 0 & 310 & 0 & 1344 \\
  -16 & 92& -32 & 386 & -64 & 1548 \\
  16 & 64 & 32 & 258 & 64 & 856 \\
  32& 16 & 64 & 50 & 128 & 288 \\
  48 & 4 & 96 & 20 & 192 & 60 \\
  \hline
\end{tabular}
\mbox{\small{Walsh spectra and the frequency of the function $f_{\lambda,1}$}}
\end{center}
We find that if we choose different $\mu$, then the frequency may be different. In other words, the frequency is related to the choice of $\mu$.

}

\subsection{Walsh spectra of $g_{\lambda,\mu}$}

In this subsection, we study the Walsh spectra of $g_{\lambda,\mu}$. The main result of this subsection is the following

{\thm Let $m\geq 1$ be a positive integer and $n=2m$. Define a Boolean function \begin{equation}\label{f-g}
    g_{\lambda,\mu}(x)=(1+{\rm Tr}_1^n(x)){\rm Tr}_1^n(\lambda x^{2^m+1})+{\rm Tr}_1^n(x){\rm Tr}_1^n(\mu x^{2^m-1}),
\end{equation}
where $\lambda$ is an element in $\mathbb{F}_{2^n}$ satisfying $\lambda+\overline{\lambda}=1$ and $\mu \in \mathbb{F}_{2^m}^*$ satisfying $k_m(\mu)=-1$. Then $g_{\lambda,\mu}$ is balanced and its Walsh spectra is at most five-valued and are contained in the set
\begin{equation*}
 \{0, \pm 2^m,\pm 2^{m+1}\}.
\end{equation*}
The nonlinearity of $g_{\lambda,\mu}$ is
\begin{equation*}
    nl(g_{\lambda,\mu})= 2^{n-1}-2^{m}.
    \end{equation*}
}

Note that the function $  g_{\lambda,\mu}$ has algebraic degree $m+1$ as well, thus it can not be bent also, but its nonlinearity is close to the maximum possible value of a Boolean function can achieve. Moreover, we shall see that this function can be balanced by a suitable choosing of the parameters $m$ and $\mu$.

\begin{proof}We prove the first statement in this subsection, the proof of the nonlinearity is left in the next subsection.

It is obvious that
\begin{equation*}
    g_{\lambda,\mu}(x)=\left\{\begin{array}{cc}
                                {\rm Tr}_1^n(\lambda x^{2^m+1}),& \mbox{ if ${\rm Tr}_1^n(x)=0$}, \\
                                {\rm Tr}_1^n(\mu x^{2^m-1}), & \mbox{ if ${\rm Tr}_1^n(x)=1$}.
                              \end{array}
    \right.
\end{equation*}
Thus we have
\begin{eqnarray*}
  W_g(0)
   &=& \sum_{x\in \mathbb{F}_{2^n}}(-1)^{g_{\lambda,\mu}(x)} \\
  &=&\sum_{x\in T_0}(-1)^{{\rm Tr}_1^n(\lambda x^{2^m+1})}+\sum_{x\in T_1}(-1)^{{\rm Tr}_1^n(\mu x^{2^m-1})}.
\end{eqnarray*}
Using the polar decomposition, we have
\begin{eqnarray*}
  &&2\sum_{x\in T_0}(-1)^{{\rm Tr}_1^n(\lambda x^{2^m+1})}\\
  &=&\sum_{x\in \mathbb{F}_{2^n}}(-1)^{{\rm Tr}_1^n(\lambda x^{2^m+1}+x)}+\sum_{x\in \mathbb{F}_{2^n}}(-1)^{{\rm Tr}_1^n(\lambda x^{2^m+1})}  \\
&=& 2+\sum_{y\in \mathbb{F}_{2^m}^*,z\in \mathfrak{S}}\Big((-1)^{{\rm Tr}_1^m(y(1+z+z^{-1}))}+(-1)^{{\rm Tr}_1^m(y)}\Big)\\
&=&\left\{\begin{array}{ll}
            2+2(2^m-1-1)+(2^m-1)(-1-1)=0, & \mbox{ if $m$ is odd}, \\
            2+(2^m+1)(-1-1)=-2^{m+1}, & \mbox{ if $m$ is even},
          \end{array}
\right.
\end{eqnarray*}
and
\begin{eqnarray*}
  &&2\sum_{x\in T_1}(-1)^{{\rm Tr}_1^n(\mu x^{2^m-1})}\\
  &=& \sum_{x\in \mathbb{F}_{2^n}}(-1)^{{\rm Tr}_1^n(\mu x^{2^m-1})}-\sum_{x\in \mathbb{F}_{2^n}}(-1)^{{\rm Tr}_1^n(\mu x^{2^m-1}+x)} \\
   &=& \sum_{y\in \mathbb{F}_{2^m}^*,z\in \mathfrak{S}}(-1)^{{\rm Tr}_1^m(\mu'(z+z^{-1}))}-\sum_{y\in \mathbb{F}_{2^m}^*,z\in \mathfrak{S}}(-1)^{{\rm Tr}_1^m(\mu'(z+z^{-1})+y(z+z^{-1}))}.
\end{eqnarray*}
By (\ref{f-kl}), we know that
\begin{equation*}
   \sum_{y\in \mathbb{F}_{2^m}^*,z\in \mathfrak{S}}(-1)^{{\rm Tr}_1^m(\mu'(z+z^{-1}))}=-(2^m-1)k_m(\mu),
\end{equation*}
and
\begin{eqnarray*}
  &&\sum_{y\in \mathbb{F}_{2^m}^*,z\in \mathfrak{S}}(-1)^{{\rm Tr}_1^m(\mu'(z+z^{-1})+y(z+z^{-1}))}\\
   &=& (2^m-1)+(-1)\sum_{z\in \mathfrak{S}\setminus \{1\}}(-1)^{{\rm Tr}_1^m(\mu'(z+z^{-1}))} \\
   &=& 2^m+k_m(\mu).
\end{eqnarray*}
Thus,
\begin{equation}\label{f-703}
    \sum_{x\in T_1}(-1)^{{\rm Tr}_1^n(\mu x^{2^m-1})}=-2^{m-1}(1+k_m(\mu)),
\end{equation}
and
\begin{equation*}
    W_g(0)=\left\{\begin{array}{cc}
                    -2^{m-1}(1+k_m(\mu)),& \mbox{ if $m$ is odd},\\
                    -2^{m-1}(3+k_m(\mu)),& \mbox{ if $m$ is even}.
                  \end{array}
    \right.
\end{equation*}
Hence, when $m$ is odd and $k_m(\mu)=-1$, then $g_{\lambda,\mu}$ is balanced; if $m$ is even, $g_{\lambda,\mu}$ is not balanced.

For any $a\in \mathbb{F}_{2^n}^*$, we have
\begin{eqnarray*}
  W_{g}(a)
  &=& \sum_{x\in \mathbb{F}_{2^n}}(-1)^{g_{\lambda,\mu}(x)+{\rm Tr}_1^n(ax)} \\
  &=& \sum_{x\in T_0}(-1)^{{\rm Tr}_1^n(\lambda x^{2^m+1}+ax)}+\sum_{x\in T_1}(-1)^{{\rm Tr}_1^n(\mu x^{2^m-1}+ax)}.
\end{eqnarray*}
The first summation is $S_0(a)$ in (\ref{f-3}). For the second summation, we have
\begin{eqnarray*}
  &&2\sum_{x\in T_1}(-1)^{{\rm Tr}_1^n(\mu x^{2^m-1}+ax)}\\
  &=& \sum_{x\in \mathbb{F}_{2^n}^*}(-1)^{{\rm Tr}_1^n(\mu x^{2^m-1}+ax)}-\sum_{x\in \mathbb{F}_{2^n}^*}(-1)^{{\rm Tr}_1^n(\mu x^{2^m-1}+(a+1)x)},
\end{eqnarray*}
and
\begin{eqnarray*}
  &&\sum_{x\in \mathbb{F}_{2^n}^*}(-1)^{{\rm Tr}_1^n(\mu x^{2^m-1}+ax)}\\
  &=&\sum_{z\in \mathfrak{S}}(-1)^{{\rm Tr}_1^m(\mu(z+z^{-1}))}\sum_{y\in \mathbb{F}_{2^m}^*}(-1)^{{\rm Tr}_1^m(y(az+\overline{a}z^{-1}))}  \\
  &=& (-1)^{{\rm Tr}_1^m(\mu^2(\frac{\overline{a}}{a}+\frac{a}{\overline{a}})}(2^m-1)+\sum_{z\in \mathfrak{S}\setminus \{(\overline{a}/a)^{1/2}\}}(-1)^{{\rm Tr}_1^m(\mu(z+z^{-1}))}(-1)\\
  &=&2^m(-1)^{{\rm Tr}_1^n(\mu^2 \overline{a}/a)}+k_m(\mu).
\end{eqnarray*}
If $a\neq 1$, then similarly we have
\begin{equation*}
   \sum_{x\in \mathbb{F}_{2^n}^*}(-1)^{{\rm Tr}_1^n(\mu x^{2^m-1}+(a+1)x)} =2^m(-1)^{{\rm Tr}_1^n(\mu^2 (\overline{a}+1)/(a+1))}+k_m(\mu).
\end{equation*}
Therefore, when $a\neq 0,1$, by  (\ref{f-3}) we have
\begin{equation*}
   W_g(a)=\left\{\begin{array}{ll}
                   2^{m-1}(-2+C) ,& \mbox{if ${\rm Tr}_1^n(a)=m(\mathrm{mod} \,2)$ and ${\rm Tr}_1^m(a\overline{a})=0$}, \\
                 2^{m-1}(2+C),&  \mbox{if ${\rm Tr}_1^n(a)=m(\mathrm{mod} \,2)$ and ${\rm Tr}_1^m(a\overline{a})=1$}, \\
                   2^{m-1}C,&  \mbox{if ${\rm Tr}_1^n(a)\not=m(\mathrm{mod} \,2)$},
                 \end{array}
   \right.
\end{equation*}
where $C=(-1)^{{\rm Tr}_1^n(\mu^2 \overline{a}/a)}-(-1)^{{\rm Tr}_1^n(\mu^2 (\overline{a}+1)/(a+1))}$.

If $a=1$, then by (\ref{f-703}),
\begin{equation*}
   \sum_{x\in T_1}(-1)^{{\rm Tr}_1^n(\mu x^{2^m-1}+x)} =-\sum_{x\in T_1}(-1)^{{\rm Tr}_1^n(\mu x^{2^m-1})}=2^{m-1}(1+k_m(\mu)).
\end{equation*}
Thus
\begin{equation*}
  W_g(1)=\left\{\begin{array}{cc}
                  2^{m-1}(1+k_m(\mu)), & \mbox{ if $m$ is odd}, \\
                  2^{m-1}(-1+k_m(\mu)), & \mbox{ if $m$ is even.}
                \end{array}
  \right.
\end{equation*}
Since $C\in \{0, \pm 2\}$, the desired result follows.
\end{proof}

{\rem  Taking $k_m(\mu)=-1$, we obtain the results while $m=3$ and $m=5$ and $m=7$ in the following table.
\begin{center}

$m=3$ \hskip 1.5 cm $m=5$ \hskip 1.5 cm $m=7$\\
\begin{tabular}{|c|c|c|c|c|c|}
  \hline
   $W_g(a)$ & frequency &$W_g(a)$ & frequency&$W_g(a)$ & frequency
 \\
  -16&4&-64 & 64  &-256 & 1016  \\
 -8&12& -32 & 236  &-128 & 4072  \\
0&24& 0 & 396  &0& 6072  \\
  8&20&32 & 260 &128 & 4216   \\
  16&4&64 & 68  &256 & 1008  \\
  \hline
\end{tabular}
\mbox{\small {Walsh spectra and frequency of the values of the function $g_{\lambda,\mu}$ with $k_m(\mu)=-1$} }

\end{center}}

\subsection{The proof of the nonlinearity} If we denote by $N_i$ the number of  $a\in \mathbb{F}_{2^n}$ such that $W_g(a)=2^mi$, $i=-2,-1,0,1,2$, then we get
$$\left\{\begin{array}{l}
  N_0=3N_2+3N_{-2}, \\
  N_1=2^{n-1}+2^{m-1}-3N_2-N_{-2}, \\
  N_{-1}=2^{n-1}-2^{m-1}-N_2-3N_{-2}.
\end{array}
\right.$$
In what follows, we give an explicit formula for $N_0$ in terms of some exponential sums.
We assume that $m$ is even in next sequel, while the case of $m$ is odd can be settled similarly.

It is evident that
\begin{eqnarray*}
  N_2 &=&|\{a: a\in \mathbb{F}_{2^n},{\rm Tr}^n_1(a)=0, {\rm Tr}_1^m(a\overline{a})=1\} \\
  &&\cap\{{\rm Tr}^n_1(\mu^2\overline{a}/a)=0, {\rm Tr}^n_1(\mu^2\overline{a+1}/(a+1))=1\}| \\
  N_{-2}&=& |\{a: a\in \mathbb{F}_{2^n},{\rm Tr}^n_1(a)=0, {\rm Tr}_1^m(a\overline{a})=0\}\\
   &&\cap\{{\rm Tr}^n_1(\mu^2\overline{a}/a)=1, {\rm Tr}^n_1(\mu^2\overline{a+1}/(a+1))=0\}|.
\end{eqnarray*}
Denote $(-1)^{{\rm Tr}_1^n(x)}$ by $\chi_n(x)$ and $(-1)^{{\rm Tr}_1^m(x)}$ by $\chi_m(x)$. Then we have
\begin{eqnarray*}
  16N_2&=&\sum_{a\in \mathbb{F}_{2^n}\setminus{\mathbb{F}_2}}(1+\chi_n(a))(1-\chi_m(a\overline{a}))(1+\chi_n(\mu^2\overline{a}/a))(1-\chi_n(\mu^2\overline{a+1}/(a+1))  \\
  16N_{-2}&=&\sum_{a\in \mathbb{F}_{2^n}\setminus{\mathbb{F}_2}}(1+\chi_n(a))(1+\chi_m(a\overline{a}))(1-\chi_n(\mu^2\overline{a}/a))(1+\chi_n(\mu^2\overline{a+1}/(a+1))  \end{eqnarray*}
Note that the term in the right hand side of the equation is zero when $\chi_n(a)=-1$. Thus substituting $a$ by $a+1$ in the second equation leads to
\begin{equation*}
     16N_{-2}=\sum_{a\in \mathbb{F}_{2^n}\setminus{\mathbb{F}_2}}(1+\chi_n(a))(1+\chi_m(a\overline{a}))(1+\chi_n(\mu^2\overline{a}/a))(1-\chi_n(\mu^2\overline{a+1}/(a+1))
\end{equation*}
Hence, it follows that
\begin{eqnarray*}
  &&8(N_2+N_{-2})\\
   &=& \sum_{a\in \mathbb{F}_{2^n}\setminus{\mathbb{F}_2}}(1+\chi_n(a))(1+\chi_n(\mu^2\overline{a}/a))(1-\chi_n(\mu^2\overline{a+1}/(a+1))\\
   &&(\mbox{substituting $a$ by $a+1$ leads to})\\
   &=& \sum_{a\in \mathbb{F}_{2^n}\setminus{\mathbb{F}_2}}(1+\chi_n(a))(1-\chi_n(\mu^2\overline{a}/a))(1+\chi_n(\mu^2\overline{a+1}/(a+1))\\
&&(\mbox{adding the above two equations and dividing by $2$ leads to})\\
&=&\sum_{a\in \mathbb{F}_{2^n}\setminus{\mathbb{F}_2}}(1+\chi_n(a))(1-\chi_n(\mu^2(\overline{a}/a+\overline{a+1}/(a+1))))\\
&=&\sum_{a\in \mathbb{F}_{2^n}\setminus{\mathbb{F}_2}}(1+\chi_n(a))(1-\chi_n(\mu^2\frac{\overline{a}+a}{a^2+a})).
\end{eqnarray*}
Therefore, we have that
\begin{eqnarray}\label{f-10242}
    N_0&=&\frac{3}{8}\sum_{a\in \mathbb{F}_{2^n}\setminus{\mathbb{F}_2}}(1+\chi_n(a))(1-\chi_n(\mu^2\frac{\overline{a}+a}{a^2+a}))\\
   \nonumber &=&\frac{3}{8}[2^n-2-2-\sum_{a\in \mathbb{F}_{2^n}\setminus{\mathbb{F}_2}}\chi_n(\mu^2\frac{\overline{a}+a}{a^2+a})-\chi_n(a+\mu^2\frac{\overline{a}+a}{a^2+a})].
\end{eqnarray}
For the sum $\sum_{a\in \mathbb{F}_{2^n}\setminus{\mathbb{F}_2}}\chi_n(\mu^2\frac{\overline{a}+a}{a^2+a})$, we have the following result which has its own independent meaning.
{\thm \label{thm10-8}
Let $\mu\in \mathbb{F}_{2^m}^*$. Then
\begin{equation}\label{f-1015}
    \sum_{a\in \mathbb{F}_{2^n}\setminus \mathbb{F}_2}\chi_n\left(\mu\frac{a^{2^m}+a}{a^2+a}\right)=-2-(1+k_m(\mu))^2.
\end{equation}

\begin{proof}In order to prove the theorem, we need to use another decomposition of elements in $\mathbb{F}_{2^n}$ instead of the polar decomposition. Denote $$E=\{\lambda\in \mathbb{F}_{2^n}: \overline{\lambda}+\lambda=1\}.$$
Then $E$ is an affine subspace of $\mathbb{F}_{2^n}/\mathbb{F}_2$. For every $x\in \mathbb{F}_{2^n}^*\setminus \mathbb{F}_{2^m}$, there is a unique pair $(u,\lambda)\in \mathbb{F}^*_{2^m}\times E$ such that $x=u\lambda$. If $x\in \mathbb{F}_{2^m}$, we just write $x=u$. This decomposition is unique, for if there are $u_1,u_2\in \mathbb{F}_{2^m}^*$ and $\lambda_1,\lambda_2\in E$ satisfying $u_1\lambda_1=u_2\lambda_2$, then
\begin{equation*}
    1=\overline{\lambda_1}+\lambda_1=(u_2/u_1)(\overline{\lambda_2}+\lambda_2)=u_2/u_1
\end{equation*}
which implies that $u_1=u_2$ and $\lambda_1=\lambda_2$. Counting the number of elements of the related sets leads to the claim.

We have the following more facts about the new decomposition of elements in $\mathbb{F}_{2^n}$

{\bf Fact }(1) For every $x\in \mathbb{F}_{2^n}\setminus \mathbb{F}_{2^m}$, let $x=u\lambda$, $u\in \mathbb{F}_{2^m}^*, \lambda\in E$. Then ${\rm Tr}_1^n(x)={\rm Tr}_1^m(u)$.

This Fact follows with the transitivity of the trace maps.

{\bf Fact }(2) The map $\sigma: E \rightarrow \mathbb{F}_{2^m}; \lambda\mapsto \lambda\overline{\lambda}$ is a two-to-one map, the image set is precisely the set of elements in $\mathbb{F}_{2^m}$ which is of trace one.

Below we give a explanation of Fact (2): It is obvious that there are two elements $\lambda_1,\lambda_2\in E$ satisfying $\lambda_1\overline{\lambda_1}=\lambda_2\overline{\lambda_2}=a$ for some $a\in \mathbb{F}_{2^m}$ if and only if $\lambda_1,\lambda_2$ are the two distinct roots of the equation
\begin{equation}\label{f-1210}
   X^2+X+a=0.
\end{equation}
And (\ref{f-1210}) has two distinct roots in $\mathbb{F}_{2^n}\setminus \mathbb{F}_{2^m}$ if and only if ${\rm Tr}_1^m(a)=1$. Moreover, for any $a\in \mathbb{F}_{2^m}$ with trace one, $\lambda$ is a root of the equation (\ref{f-1210}) if and only if $\overline{\lambda}$ is also a root of the equation, thus $\lambda+\overline{\lambda}=1$ and $\lambda\in E$. This proves the Fact (2).

Since
\begin{eqnarray*}
   \sum_{a\in \mathbb{F}_{2^n}\setminus{\mathbb{F}_2}}\chi_n(\mu^2\frac{\overline{a}+a}{a^2+a})&=&\sum_{a\in \mathbb{F}_{2^m}\setminus{\mathbb{F}_2}}1+\sum_{a\in \mathbb{F}_{2^n}\setminus \mathbb{F}_{2^m}}\chi_n(\mu^2\frac{\overline{a}+a}{a^2+a}) \\
   &=&2^m-2+ \sum_{a\in \mathbb{F}_{2^n}\setminus \mathbb{F}_{2^m}}\chi_n(\mu^2\frac{\overline{a}+a}{a^2+a}).
\end{eqnarray*}
 By the Fact (2), one has that
\begin{eqnarray*}
  &&\sum_{u\in \mathbb{F}_{2^m}^*,\lambda\in E}\chi_m\left(\mu (\frac{1}{\overline{\lambda}\lambda}+\frac{1}{\lambda\overline{\lambda}+u^{2}+u})\right)\\
   &=&2\sum_{v\in \mathbb{F}_{2^m}, {\rm Tr}_1^m(v)=1}\chi_m(\frac{\mu }{v})\sum_{u\in \mathbb{F}_{2^m}^*}\chi_m\left(\frac{\mu}{v+u^2+u}\right)\\
  &=&2\sum_{v\in \mathbb{F}_{2^m}, {\rm Tr}_1^m(v)=1}\chi_m(\frac{\mu }{v})\left(\chi_m(\mu/v)+2\sum_{u\in \mathbb{F}_{2^m}^*\setminus\{v\}, {\rm Tr}_1^m(u)=1}\chi_m(\frac{\mu}{u})\right).
\end{eqnarray*}
Since
\begin{eqnarray*}
 &&\chi_m(\mu/v)+2\sum_{u\in \mathbb{F}_{2^m}^*\setminus\{v\}, {\rm Tr}_1^m(u)=1}\chi_m(\frac{\mu}{u}) \\
   &=&\chi_m(\mu/v)-2\chi_m(\mu/v)+2\sum_{u\in \mathbb{F}_{2^m}, {\rm Tr}_1^m(u)=1}\chi_m(\mu/u)\\
 &=&\chi_m(\mu/v)-2\chi_m(\mu/v)+\sum_{u\in \mathbb{F}_{2^m}}\chi_m(\mu/u)(1-\chi_m(u))
 \\
   &=&-\chi_m(\mu/v)-\sum_{u\in \mathbb{F}_{2^m}}\chi_m(u+\mu/u)\\
 &=&-1-\chi_m(\mu/v)-k_m(\mu),
\end{eqnarray*}
one has that
\begin{eqnarray*}
 &&\sum_{a\in \mathbb{F}_{2^n}\setminus \mathbb{F}_{2^m}}\chi_n\left(\mu\frac{a^{2^m}+a}{a^2+a}\right)\\
   &=& -2(1+k_m(\mu))\sum_{v\in \mathbb{F}_{2^m}, {\rm Tr}_1^m(v)=1}\chi_m(\mu/v)-2\sum_{v\in \mathbb{F}_{2^m}, {\rm Tr}_1^m(v)=1}1\\
   &=&-(1+k_m(\mu))^2-2^m.
\end{eqnarray*}
Therefore, we have
\begin{equation}\label{f-4}
 \sum_{a\in \mathbb{F}_{2^n}\setminus \mathbb{F}_2}\chi_n\left(\mu\frac{a^{2^m}+a}{a^2+a}\right)=-2-(1+k_m(\mu))^2.
\end{equation}
 This completes the proof.
\end{proof}
}



For the sum $  \sum_{a\in \mathbb{F}_{2^n}\setminus{\mathbb{F}_2}}\chi_n(a+\mu^2\frac{\overline{a}+a}{a^2+a})$, we compute
\begin{eqnarray*}
   \sum_{a\in \mathbb{F}_{2^n}\setminus{\mathbb{F}_2}}\chi_n(a+\mu^2\frac{\overline{a}+a}{a^2+a})&=&\sum_{a\in \mathbb{F}_{2^m}\setminus{\mathbb{F}_2}}1+\sum_{a\in \mathbb{F}_{2^n}\setminus \mathbb{F}_{2^m}}\chi_n(a+\mu^2\frac{\overline{a}+a}{a^2+a}) \\
   &=&2^m-2+ \sum_{a\in \mathbb{F}_{2^n}\setminus \mathbb{F}_{2^m}}\chi_n(a+\mu^2\frac{\overline{a}+a}{a^2+a}).
\end{eqnarray*}
And
\begin{eqnarray*}
  &&\sum_{a\in \mathbb{F}_{2^n}\setminus \mathbb{F}_{2^m}}\chi_n(a+\mu^2\frac{\overline{a}+a}{a^2+a})\\
   &=&\sum_{a\in \mathbb{F}_{2^n}\setminus \mathbb{F}_{2^m}}\chi_n(\mu^2\frac{\overline{a}+a}{a^2+a})-2\sum_{a\in \mathbb{F}_{2^n}\setminus \mathbb{F}_{2^m}, {\rm Tr}_1^n(a)=1}\chi_n(\mu^2\frac{\overline{a}+a}{a^2+a})  \\
 &=& -2^m-2\sum_{a\in \mathbb{F}_{2^n}\setminus \mathbb{F}_{2^m}, {\rm Tr}_1^n(a)=1}\chi_n(\mu^2\frac{\overline{a}+a}{a^2+a})\\
 &=&-2^m-2\sum_{{\rm Tr}_1^m(u)=1, \lambda\in E}\chi_n(\mu^2(\frac{1}{\lambda}+\frac{1}{\lambda+1/u})).
\end{eqnarray*}

By Remark 3.7, we have
\begin{eqnarray}
  \nonumber&&2\sum_{{\rm Tr}_1^m(u)=1, \lambda\in E}\chi_n(\mu^2(\frac{1}{\lambda}+\frac{1}{\lambda+1/u}))\\
  \nonumber&=&2\sum_{{\rm Tr}_1^m(1/u)=1, \lambda\in E}\chi_n(\mu^2(\frac{1}{\lambda}+\frac{1}{\lambda+u}))  \\
  \nonumber &=&2 \sum_{{\rm Tr}_1^m(1/u)=1, \lambda\in E}\chi_m(\mu^2(\frac{1}{\lambda\overline{\lambda}}+\frac{1}{\lambda\overline{\lambda}+u^2+u}))\\
   \nonumber&=&4\sum_{{\rm Tr}_1^m(1/u)=1, {\rm Tr}_1^m(v)=1}\chi_m(\mu^2(\frac{1}{v}+\frac{1}{v+u^2+u}))\\
   \nonumber&=:&4R(u),
\end{eqnarray}
and
\begin{equation}\label{f-10301}
    \sum_{a\in \mathbb{F}_{2^n}\setminus \mathbb{F}_{2^m}}\chi_n(a+\mu^2\frac{\overline{a}+a}{a^2+a})=-2^m-4R(\mu).
\end{equation}

Therefore, by (\ref{f-10242}) and (\ref{f-1015}), (\ref{f-10301}), we get that
\begin{equation*}
    N_0=\frac{3}{8}(2^n+4R(u))=\frac{3}{2}(2^{n-2}+R(u)).
\end{equation*}
We claim that there is at least one pair of $(u,v)$ with ${\rm Tr}_1^m(1/u)=1, {\rm Tr}_1^m(v)=1$ such that $\chi_m(\mu^2(\frac{1}{v}+\frac{1}{v+u^2+u}))=1$. We use reduction to absurdity to prove this claim. If for every pair of $(u,v)$ with ${\rm Tr}_1^m(1/u)=1, {\rm Tr}_1^m(v)=1$, one has that $\chi_m(\mu^2(\frac{1}{v}+\frac{1}{v+u^2+u}))=-1$, then taking $v=1/u$, one should have that

 \begin{equation*}
    \chi_m(\mu^2\frac{1+v^2}{v^4+v^3+v})=-1 \mbox{ for all $v$ satisfying ${\rm Tr}_1^m(v)=1$}.
\end{equation*}
In other words, it holds that for a fixed $v_0\in \mathbb{F}_{2^m}$ with ${\rm Tr}_1^m(v_0)=1$ and all $z\in \mathbb{F}_{2^m}$,
\begin{equation}\label{f-10311}
     \chi_m(\mu^2\frac{\Gamma_1(z)}{\Gamma_2(z)})=-1,
\end{equation}
where $\Gamma_1(z)=1+z^4+z^2+v_0^2$ and $\Gamma_2(z)=z^8+z^6+z^5+v_0z^4+z^3+(v_0^2+1)z^2+(v_0^2+v_0+1)z+v_0^4+v_0^3+v_0$, i.e.,
\begin{equation}\label{f-10312}
   \sum_{z\in \mathbb{F}_{2^m}} \chi_m(\mu^2\frac{\Gamma_1(z)}{\Gamma_2(z)})=-2^m.
\end{equation}
However by a well-known exponential bound obtained by C.J. Moreno and O. Moreno, see \cite{momo}, Theorem 2, we have
\begin{equation*}
   \left|\sum_{z\in \mathbb{F}_{2^m}} \chi_m(\mu^2\frac{\Gamma_1(z)}{\Gamma_2(z)})\right|\leq 14 \sqrt{2^m}+1<2^m
\end{equation*}
if $m\geq 8$. For the cases of $m< 8$, the fact of $N_0>0$ were verified by using Magma.
Therefore, the claim is proved and then $|R(\mu)|<2^{n-2}$ which leads to $N_0>0$.
Thus $g_{\lambda,\mu}$ can not be a bent function. Moreover, from $N_0>0$, one has that at least one of the numbers $N_2$ and $N_{-2}$ is nonzero, and the nonlinearity of the function is $2^{n-1}-2^m$.


\section*{Acknowledgement}

The work of this paper is supported by the NUAA Fundamental Research Funds, No. 2013202 and NNSF of China under Grant No. 11371011.


\end{document}